\documentclass[a4paper]{article}
\usepackage{graphicx}

\newcommand{\chicj}{\chi_{cJ}}

\newcommand{\chico}{\chi_{c1}}
\newcommand{\chict}{\chi_{c2}}
\newcommand{\psip}{\psi(2S)}

\newcommand{\jpsi}{J/\psi}
\newcommand{\ccbar}{c\overline{c}}

\newcommand{\ppbar}{p\overline{p}}
\newcommand{\ldldbar}{\Lambda \overline{\Lambda}}
\newcommand{\sgsgzbar}{\Sigma \overline{\Sigma}^0}

\newcommand{\EE}{e^+e^-}

\newcommand{\GG}{\gamma\gamma}
\newcommand{\PP}{\pi^+\pi^-}
\newcommand{\KK}{K^+K^-}
\newcommand{\PPJP}{\pi^+\pi^- J/\psi}
\newcommand{\kskl}{K^0_SK^0_L}

\newcommand{\ra}{\rightarrow}

\newcommand{\rhopi}{\rho\pi}

\newcommand{\beq}{\begin{equation}}
\newcommand{\eeq}{\end{equation}}
\newcommand{\beqn}{\begin{eqnarray}}
\newcommand{\eeqn}{\end{eqnarray}}
\newcommand{\beqns}{\begin{eqnarray*}}
\newcommand{\eeqns}{\end{eqnarray*}}
\newcommand{\bfg}{\begin{figure}}
\newcommand{\efg}{\end{figure}}
\newcommand{\bitm}{\begin{itemize}}
\newcommand{\eitm}{\end{itemize}}
\newcommand{\bnum}{\begin{enumerate}}
\newcommand{\enum}{\end{enumerate}}
\newcommand{\btbl}{\begin{table}}
\newcommand{\etbl}{\end{table}}
\newcommand{\btbu}{\begin{tabular}}
\newcommand{\etbu}{\end{tabular}}
\begin{document}
\normalsize

\title{Recent Results of $\psip$ Decays at BES}
\author{XiaoHu Mo \\
{\small \em Institute of High Energy Physics, CAS,
Beijing 100039, China} }
\date{}
\maketitle
\begin{abstract}
Using 14 million $\psip$ data sample collected with BES at BEPC,
$\psip \rightarrow VT$, $\kskl$ (also $\jpsi \ra \kskl$), 
and $\chicj \ra B\overline{B}$ decays are measured and compared with
theoretical model predications.
\end{abstract}

Charmonium physics is one of the interesting and intriguing field of
particle physics. Charmonium provides us an excellent and simple system to
study QCD, the production and decay mechanisms of heavy quarkonia and light
hadron spectra from its decays, and can be treated non-relativistically. 
Using 14 M $\psip$ data sample collected with BEijing
Spectrometer (BES) at BEPC, $\psip \rightarrow VT$, $\kskl$ 
(also $\jpsi \ra \kskl$), and $\chicj \ra B\overline{B}$ decays are
measured and compared with theoretical model predications.
The BES detector is described in detail in Ref.\cite{besdct}.

\section{Study of VT Channel in $\psip$ Decay}
\begin{figure}[hbt]
\begin{minipage}{7cm}
\includegraphics[height=6.cm,width=7.cm,angle=0]{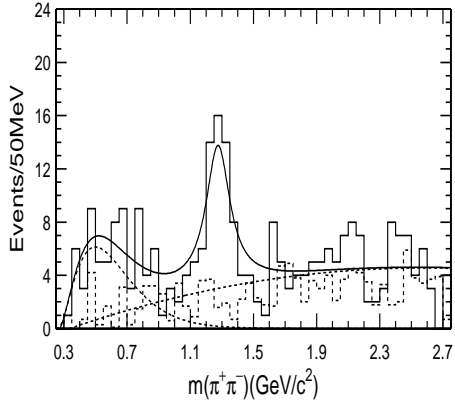}
\begin{center} (a) $\omega f_2$ final state \end{center}
\end{minipage}
\begin{minipage}{7cm}
\includegraphics[height=6.cm,width=7.cm,angle=0]{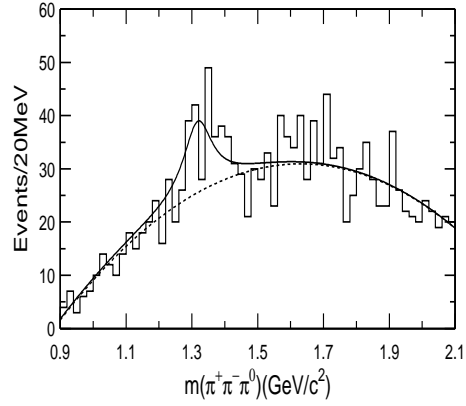}
\begin{center} (b) $\rho a_2$ final state \end{center}
\end{minipage}
\begin{minipage}{7cm}
\includegraphics[height=6.cm,width=7.cm,angle=0]{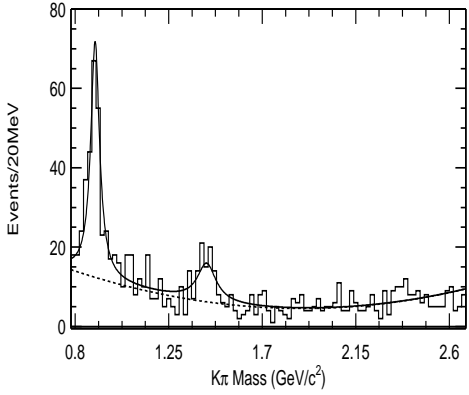}
\begin{center} (c) $K^* \overline{K^*_2}$ final state  \end{center}
\end{minipage}
\begin{minipage}{7cm}
\includegraphics[height=6.cm,width=7.cm,angle=0]{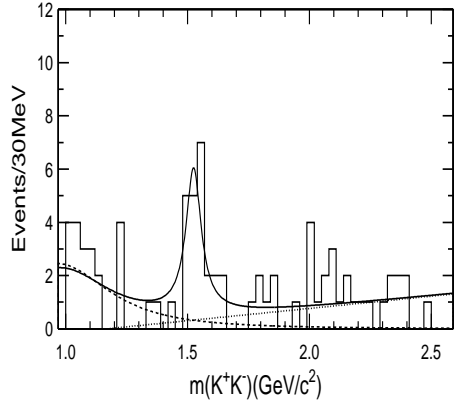}
\begin{center} (d) $\phi f_2$ final state \end{center}
\end{minipage}
\caption{\label{vtfig} Invariant mass distributions for VT channel(The dashed
line indicates background and solid line indicates the synthetic fitting 
result).}
\end{figure}

Both $\jpsi$ and $\psip$ decays are expected to be dominated by annihilation
into three gluons, with widths that are proportional to the square of the 
$\ccbar$ wave function at the origin~\cite{appelquist}. This yields the 
pQCD expectation (so-called ``12 \% '' rule) that
\beq
Q_h =\frac{{\cal B}_{\psip \ra X_h}}{{\cal B}_{\jpsi \ra X_h}}
=\frac{{\cal B}_{\psip \ra \EE}}{{\cal B}_{\jpsi \ra \EE}} 
= (12.3 \pm 0.7) \%~~.
\label{pqcdrule}
\eeq
 The violation of this rule was firstly revealed by MARK-I in VP channel (such
 as $\rhopi$ and $K^* \overline{K}$ channel)~\cite{mk2}, which leads to famous 
 ``$\rhopi ~puzzle$''. This phenomenon was then confirmed by BES at higher 
 sensitivity~\cite{zhuys}. Afterwards, BES collaboration presented many other 
 observations, one of them is about VT channel. Based on BES-I 4 M data, 
 the upper-limits of four VT channels, $\omega f_2(1270)$, $\rho a_2(1320)$, 
 $K^*(892)^0 \overline{K^*_2}(1430)^0~+c.c.$, and $\phi f_2^{\prime}(1525)$, 
 were given~\cite{vtbes1}. Now with BES-II 14 M $\psip$ date sample, 
 all these upper-limits have been determined to be
 branching fractions. For these decay modes, 
studies focus on four-charged-track final states, such as $\KK\PP$ or $\KK\KK$, 
and those with additional two photons decayed from $\pi^0$, such as $\PP\PP \GG$ or $\KK \PP \GG$. After event selection, the invariant mass distributions for 
different channels are shown in Fig.~\ref{vtfig}. From the data fitting, the 
observed numbers of events are obtained, and M.C. simulation gives the 
corresponding efficiencies. The final results are shown in 
Table~\ref{reslist}, together with statistical and systematic errors, and
the resulting errors are at the level of 30 to 40 percent. Combining the 
corresponding results of $\jpsi$ decay from PDG2002~\cite{pdg}, the $Q_h$ values were calculated as listed in Table~\ref{reslist}. Comparing with 12 \% rule, it is seen that the $Q_h$ value of VT channel is greatly suppressed.

\section{First Observation of $\kskl$ in $\psip$ Decay}
\begin{figure}
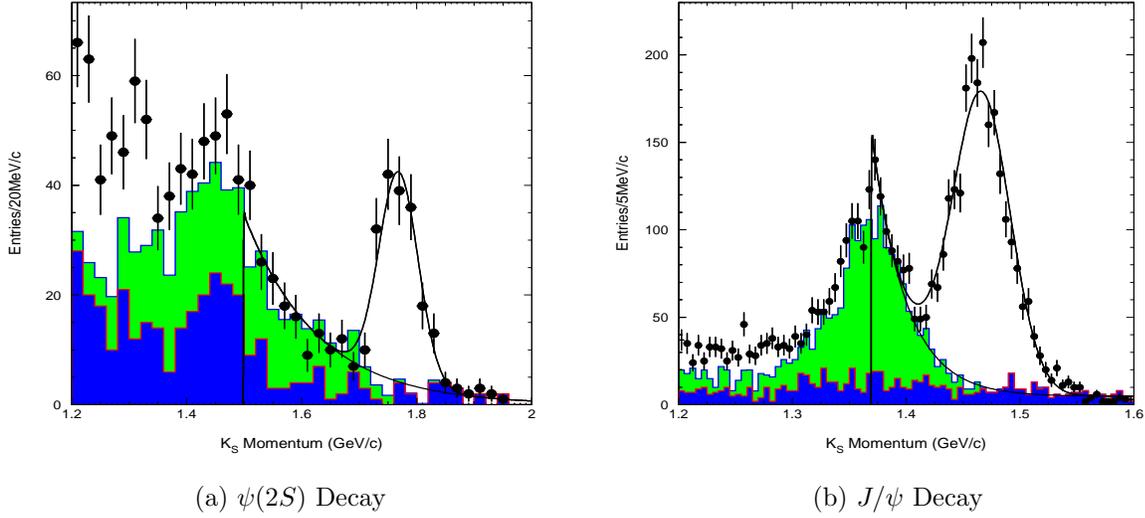

\begin{minipage}{7.5cm}
\includegraphics[height=6.cm,width=7.cm,angle=0]{pksdtfit_prl.epsi}
\center (a) $\psip$ Decay
\end{minipage}
\hskip 0.5cm
\begin{minipage}{7.5cm}
\includegraphics[height=6.0cm,width=7.cm,angle=0]{pksfit_j.epsi}
\center (b) $\jpsi$ Decay
\end{minipage}
\caption{The $K^0_S$ momentum distribution for (a) $\psip$ decay and 
(b) $\jpsi$ decay. The dots with error bars are data, the dark
shaded histogram is from $K^0_S$ mass sideband events, and the light shaded 
histogram is Monte Carlo simulated backgrounds. The curves shown in the plot
are from a best fit of the distribution. }
\label{psipjkskl}
\end{figure}

For pseudoscalar meson pairs (PP) decays, the theoretical motivation, 
besides the pCQD rule test, involves the phase study which is very important in understanding the strong
interaction mechanism of charmonium decay. A recent phenomenological analysis 
predicts a relation between the branching ratio of $\kskl$ and the phase 
between the three-gluon and the one-photon annihilation amplitudes~\cite{phase}. So the measurement of the $\kskl$ branching ratio is important to determine the phase. From data analysis point of view, the event topology of 
$\psip \ra \kskl$ is fairly prominent: the neutral $K_L$ almost leaves no 
information in Main Drift Chamber due to long decay lifetime, while the $K_S$ 
swiftly decays into two pions. By the virtue of this characteristic topology of event, two good charged tracks are required with net charge zero; in addition, 
secondary vertex requirement is applied for $K_S$ identification. With these 
requirements, the distribution of 
the momentum of $K_S$ is obtained as shown in Fig.~\ref{psipjkskl}~(a). 
The different shaded histograms indicate different estimation and simulation of background, whose shape in the vicinity of signal region is described by 
exponential function. For signal 
events, the Gaussian function is used to fit the observed number of events. The final branching ratio is worked out to be 
$(5.25 \pm 0.47 \pm 0.63) \times 10^{-5}$. 
The similar study has also been made for $\jpsi \rightarrow \kskl$ decay. 
The momentum distribution of $K_S$ is shown in Fig.~\ref{psipjkskl}~(b) and the 
branching ratio is worked out to be $(1.86\pm 0.47 \pm 0.63) \times 10^{-4}$. 
It is worth while to notice that the BES measurement result is considerably 
larger than that of the PDG value: ${\cal B}_{\jpsi \ra \kskl}=
(1.08\pm 0.47) \times 10^{-4}$.

\begin{table}
\begin{center}
\caption{\label{reslist} The results of $\psip$ and $\chicj$ decays.}
\begin{tabular}{cccc} \hline \hline
 VT channel &  ${\cal B}_{\psip}~(10^{-4})$ 
            &  ${\cal B}_{\jpsi}~(10^{-3})$ & $Q_h$ (\%) \\ 
            &  (from BES)       
	    &  (from PDG2002)                &       \\ \hline
$\omega f_2$& $2.05 \pm 0.41 \pm 0.46$  
            & $4.3  \pm 0.6          $       & $4.8 \pm 1.5$ \\
$\rho a_2$  & $2.55 \pm 0.73 \pm 0.60$    
            & $10.9 \pm 2.2          $       & $2.3 \pm 1.1$ \\
$K^* \overline{K^*_2} + c.c.$
            & $1.64 \pm 0.33 \pm 0.41$    
            & $6.7  \pm 2.6          $       & $2.4 \pm 1.2$ \\
$\phi f_2^{\prime}$  
            & $0.48 \pm 0.14 \pm 0.12$
	    & $1.23 \pm 0.06 \pm 0.20$ $\ast$
	                                     & $3.9 \pm 1.6$ \\  \hline \hline
 PP channel & ${\cal B}_{\psip}~(10^{-5})$ 
            & ${\cal B}_{\jpsi}~(10^{-4})$   & $Q_h$ (\%) \\ 
            &  (from BES)         
	    &  (from BES)                    &       \\ \hline
 $\kskl$    & $5.25 \pm 0.47 \pm 0.63$
            & $1.86 \pm 0.43 \pm 1.2$        & $28.2 \pm 4.7$ \\ \hline  \hline
Decay mode  & ${\cal B}_{Exp.}~(10^{-4})$
            & ${\cal B}_{The.}~(10^{-4})$   & $R_{Exp./The.}$ \\
            &  (from BES)         & (by COM)           &       \\ \hline
$\chi_{c0} \ra \ldldbar$ 
            & $4.7^{+1.3}_{-1.2} \pm 1.0 $  
	    & $-$      & $-$  \\
$\chi_{c1} \ra \ldldbar$ 
            & $2.6^{+1.0}_{-0.9} \pm 0.6 $ 
	    & 0.366    & 7.1  \\
$\chi_{c2} \ra \ldldbar$ 
            & $3.3^{+1.5}_{-1.3} \pm 0.7 $
	    & 0.333    & 9.9  \\ \hline \hline
\end{tabular}\\
{\small $\ast$ This value from DM2 only.} 
\end{center}
\end{table}

\begin{figure}
\begin{minipage}{7.5cm}
\includegraphics[height=6.0cm,width=7.cm,angle=0]{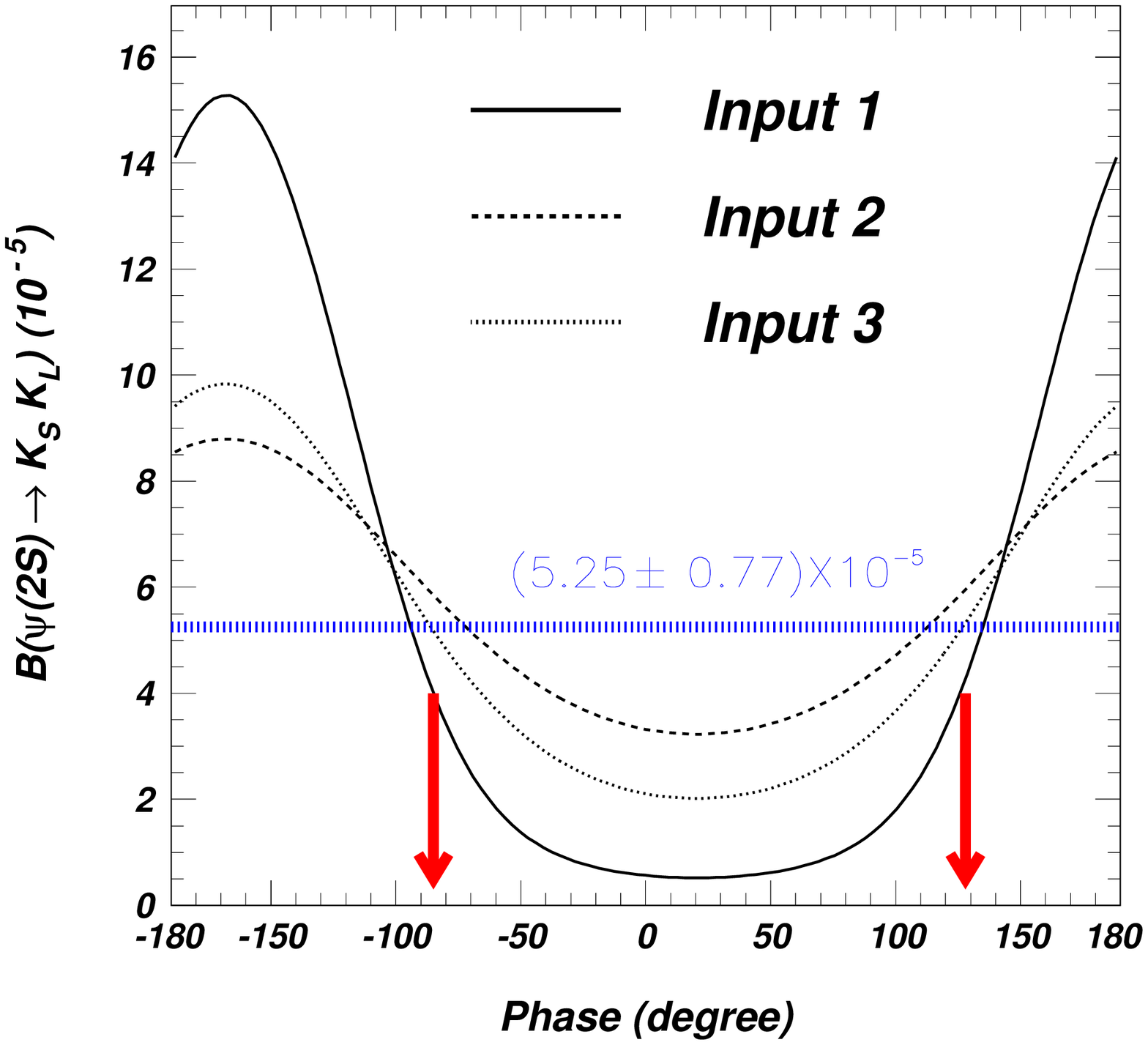}
\center (a) Relation between ${\cal B}(\kskl)$ and phase
\end{minipage}
\hskip 0.5cm
\begin{minipage}{7.5cm}
\includegraphics[height=6.0cm,width=7.cm,angle=0]{maa-fit.epsi}
\center (b) Mass distribution of $\gamma \ldldbar$ 
\end{minipage}
\caption{(a) $\psip \ra \kskl$ branching ratio as a function of the relative 
phase for three different inputs: input one based on DASP results; 
input two on BES results;
input three on $\KK$ result from BES and $\PP$ result derived from pion form
factor (for detail information, see Ref.~\cite{phase}).
(b) Mass distribution of $\gamma \ldldbar$ candidates, 
fit with three mass resolution smeared Breit-Wigner functions and a background
as estimated from data sideband and Monte Carlo simulation.}
\label{phiandm}
\end{figure}

In contrast with VT channel, the $Q_h$ value for $\kskl$ channel is enhanced greatly. Using BES measured branching ratios of $\kskl$ decay from $\jpsi$ and $\psip$,
the $Q_h$ value is calculated as ($28.2 \pm 4.7$)\%.
Comparing to 12\% rule, the deviation is greater than 3 $\sigma$.
According to Ref.~\cite{phase}, the relation between branching
ratio of $\psip \rightarrow \kskl$ and the phase between the three-gluon and the one-photon annihilation amplitudes is shown in Fig.~\ref{phiandm}(a) , where 
three inputs correspond to three groups of branching ratios of $\psip$ to $\PP$ or $\KK$. Using the $\kskl$ branching ratio, the phase is determined to be
either $-85^{\circ}$ or $130^{\circ}$. Here the most interesting result is
the large phase which supports the theoretically favored orthogonal phase
assumption~\cite{gerard}.

\section{Study of $\chicj$ Decays}
The large sample of $\psip$ decays permits studies of $\chicj$ decays with
high precision. Some theoretical papers of interest are given in 
Ref.~\cite{BBref}, and experiment results from BES could refer to 
Refs.~\cite{chires} and~\cite{chisum}. A recent analysis involving $\chicj$ 
decay is the measurement of branching ratios of $\chicj \ra \ldldbar$. The detailed information could be
found in Ref.~\cite{chicjldld}, where $\gamma \PP \ppbar$ events with 
$\PP \ppbar$ mass in the $\chicj$ mass region are studied carefully. The 
background from non-$\ldldbar$ event is estimated from the $\Lambda$ mass
sidebands of data distribution, while that from channels with $\ldldbar$ 
production is estimated according to Monte Carlo simulation for the following
decay modes: $\psip \ra \ldldbar, \sgsgzbar, \Lambda \overline{\Sigma}^0+c.c.~,
\Xi \overline{\Sigma}^0+c.c.~,$ and $\psip \ra \gamma \chicj,~ \chicj \ra 
\sgsgzbar \ra \GG \ldldbar$. In addition, $\psip \ra \PPJP \ra \PP \ppbar$
as background is also taken into consideration. The background shape is
determined by combining two kinds of background estimation, and the observed
numbers of events are obtained from fitting of the selected $\ldldbar$ mass
spectrum, as shown in Fig.~\ref{phiandm}(b), and the branching ratios could be found in Table~\ref{reslist}.

For comparison, the relevant theoretical results are also listed in 
Table~\ref{reslist}, where the theoretical calculation is based on Color Octet
Mechanism (COM). According to the values listed in the table, it could be seen the 
results on $\chico$ and $\chict$ decays only agree marginally with model 
predictions.

\section{Summary}
 Based on 14 M $\psip$ data sample, the branching ratios of four VT channels,
 $\omega f_2$, $\rho a_2$,$K^* \overline{K^*_2}+c.c.$, and $\phi f_2^{\prime}$
 are measured. The suppression of this decay mode with respect to ``12\%'' rule
 is confirmed with better precision. The final state $\kskl$ is first observed in $\psip $ decay. The $Q_h$ of this final state is calculated and
 is considerably enhanced comparing with ``12\%'' rule.
 In addition, using the branching ratio of $\psip \ra \kskl$, the phase
 between the three-gluon and the one-photon annihilation
 amplitudes is determined to be either $-85^{\circ}$ or $130^{\circ}$.
 The branching ratios of $\chicj \rightarrow \ldldbar$ are measured,
 with these results, the effectiveness of the calculation based on Color
 Octet Mechanism was tested.

\section*{Acknowledgment}
I would like to acknowledge help from Prof. C.~Z. Yuan and Dr. W.~F. Wang, who provided corresponding results presented in this report.
Thanks are also due to Prof. Y.~S. Zhu, Prof. S.~Jin, Prof. C.~Z. Yuan, Dr. W.~F. Wang and other colleagues of BES collaboration who gave many suggestions and comments for improvement of my report.


\begin{thebibliography}{99}
\bibitem{besdct}J.~Z.~Bai, {\em et al.}, (BES Collab.), 
Nucl. Inst. Meth. {\bf A344}, 319~(1994);{\bf A458}, 627~(2001).
\bibitem{appelquist}T.~Appelquist and H.~D.~Politzer, 
Phys. Rev. Lett. {\bf 34}, 43~(1975);\\
A.~De R\'{u}jula and S.~L.~Glashow,
Phys. Rev. Lett. {\bf 34}, 46~(1975).
\bibitem{mk2}M.~E.~B.~Franklin {\em et al.}, (MARKII Collab.),
Phys. Rev. Lett. {\bf 51}, 963~(1983).
\bibitem{zhuys}Y.~S.~Zhu, Proc. of the 28th Intern. Conf. on High Energy 
Physics (Warsaw University, 1996), Eds. Z.~Ajduk and A.~K.~Wroblewski (World
Scientific, 1997) p.507.
\bibitem{vtbes1}J.~Z.~Bai, {\em et al.}, (BES Collab.),  
Phys. Rev. Lett. {\bf 81}, 5080~(1998).
\bibitem{pdg}Particle Data Group, K.~Hagiwara {\em et al.},
Phys. Rev. {\bf D66}, 01001 (2002).       
\bibitem{phase}C.~Z.~yuan, P.~Wang and X.~H.~Mo, 
Phys. Lett. {\bf B567}, 73~(2003). 
\bibitem{gerard}J.-M.~G\'{e}rard and J.~Weyers,
Phys. Lett. {\bf B462}, 324 (1999).
\bibitem{BBref}J.~Bolz, P.~Kroll and G.~A.~Schuler, 
Phys. Lett. {\bf B392}, 198~(1997);
G.~T.~Bodwin, E.~Braaten and G.~P.~Lepage, 
Phys. Rev. {\bf D51}, 1125~(1995);
P.~Kroll and S.~M.~H.~Wong, hep-ph/9710464;
V.~L.~Chernyak and A.~R.~Zhitnitsky, 
Nucl. Phys. {\bf B201}, 492~(1982);
S.~J.~Brodsky and G.~P.~Lepage, 
Phys. Rev. {\bf D24}, 2848~(1995);
A.~Duncan and A.~H.~Mueller, 
Phys. Lett. {\bf B93}, 119~(1980).
\bibitem{chires}J.~Z.~Bai, {\em et al.}, (BES Collab.),
Phys. Rev. Lett. {\bf 81}, 3091~(1998).
\bibitem{chisum}F.~Liu,
Nucl. Phys. {\bf A675}, 71c-75c~(2000).
\bibitem{chicjldld}J.~Z.~Bai, {\em et al.}, (BES Collab.),
Phys. Rev. {\bf D67}, 112001~(2003).
\end{thebibliography}
\end{document}